\documentclass[aps,prc,letterpaper,11pt,twoside,tightenlines,nofootinbib,showpacs,preprint]{revtex4-1}
\usepackage[latin1]{inputenc}
\usepackage{amsmath}
\usepackage{amsfonts}
\usepackage{amssymb}
\usepackage{enumerate}
\usepackage{graphicx}
\usepackage{subfigure}
\usepackage{rotating}
\begin{document}
\title{Reanalysis of the $e^+ e^- \to (D^* \bar D^*)^{\pm} \pi^{\mp}$ reaction and the claim for the $Z_c (4025)$ resonance.}

\author{A. Mart\'inez Torres$^1$, K. P. Khemchandani$^1$, F. S. Navarra$^1$, M. Nielsen$^1$ and E. Oset$^{1,2}$}
\affiliation{$^1$ Instituto de F\'{\i}sica,
Universidade de Sao Paulo, C.P. 66318,
05389-970 Sao Paulo, SP, Brazil}
\affiliation{$^2$ Departamento de F\'{\i}sica Te\'orica and IFIC, Centro Mixto Universidad de Valencia-CSIC,
Institutos de Investigaci\'on de Paterna, Aptdo. 22085, 46071 Valencia,Spain}

\date{\today}

\begin{abstract} 
In this paper we study the reaction $e^+ e^- \to (D^* \bar D^*)^{\pm} \pi^{\mp}$ in which the BESIII collaboration has claimed the existence
of a $1^+$ resonance, named $Z_c(4025)$, in the $D^*\bar D^*$ invariant mass spectrum with a mass around 4026 MeV and width close to 26 MeV . We determine
the $D^*\bar D^*$ invariant mass distribution and find that although the explanation considered by the BESIII collaboration is plausible, 
there are others which are equally possible, like a $2^+$ resonance or a bound state. Even more, we find that
the data can be explained without the existence of a resonance/bound state. In view of the different possible
interpretations found for the BESIII data, we try to devise a strategy which could help in identifying the origin of the signal reported by the BESIII collaboration. For this, we study the dependence of the
$D^*\bar D^*$ spectrum considering the different options as a function of the total center of mass energy. We arrive to the conclusion that increasing
the center of mass energy from 4.26 GeV to 4.6 GeV can be useful to distinguish between a resonance, a bound state or just a pure background as being responsible
for the signal found. This information should be useful for future experiments. 
\end{abstract}


\maketitle

\section{Introduction}

In an experiment on the $e^+ e^- \to (D^* \bar D^*)^{\pm} \pi^{\mp}$ reaction at $\sqrt s = 4.26$~GeV \cite{besexp} the BESIII Collaboration reported a peak seen in the $(D^* \bar D^*)^{\pm}$ invariant mass distribution just about 10 MeV above threshold. The peak was identified as a new particle, the $Z_c(4025)$. The authors assume in the paper that the $(D^* \bar D^*)^{\pm}$ pair is created in a $S$-wave and then the $Z_c(4025)$ state, coupling to $(D^* \bar D^*)^{\pm}$, has $J^P=1^+$ to match, together with the pion, the quantum numbers $J^P=1^-$ of the virtual photon from the $e^+ e^-$ pair. However, they also state that the experiment does not exclude other spin-parity assignments. On the other hand, since the $(D^* \bar D^*)^{\pm}$ has charge, the isospin must be $I=1$.

  Bumps close to the threshold of a pair of particles are sometimes identified as new particles, but they can also be a reflection of a resonance below threshold. Indeed, in a similar reaction, $e^+ e^- \to J/\psi~ D \bar D$ \cite{Abe:2007sya}, the Belle collaboration reported a bump close to the threshold in the $D \bar D$ invariant mass distribution, which was tentatively interpreted as a new resonance. Yet, in Ref.~\cite{danibump} it was shown that the bump was better interpreted in terms of a $D \bar D$ molecular state, below the  $D \bar D$ threshold (so called X(3700)), that had been predicted in Ref.~\cite{daniddbar} and has also been found theoretically in other works 
\cite{HidalgoDuque:2012pq,Nieves:2012tt,Liu:2010xh,Zhang:2006ix,zzy1}. Similarly, in Ref.~\cite{MartinezTorres:2012du}, the $\phi \omega$ threshold peak seen in the $J/\psi \to \gamma \phi \omega$ reaction \cite{Ablikim:2006dw} was better interpreted as a signal of the $f_0(1710)$ resonance, below the $\phi \omega$ threshold, which couples strongly to $\phi \omega$~\cite{gengvec}. Another such case has been recently reported in Ref.~\cite{miguel}, where a bump close to threshold in the $K^{*0}\bar{K}^{*0}$ invariant mass distribution, seen in the $J/\psi \to \eta K^{*0}\bar{K}^{*0}$ decay in Ref.~\cite{Ablikim:2009ac}, is interpreted as a signal of the formation of an $h_1$ resonance, predicted in Ref.~\cite{gengvec}, which couples mostly to the $K^* \bar K^*$.

In the present case one investigates a peak close to threshold of $D^* \bar D^*$. Hence, the first question to ask is what we know about the interaction between this pair of mesons and whether a bound state of $D^* \bar D^*$ with isospin $I=1$, as in the present reaction, can be formed. An answer to this question can be seen in Ref.~\cite{raquelxyz}, where vector-vector hidden charm states are investigated using an interaction from an extrapolation of the local hidden gauge approach \cite{hidden1,hidden2,hidden3,hidden4}. In Ref.~\cite{raquelxyz} several states are found which can be matched with some X,Y,Z states reported in the literature, and there is only one bound state of $D^* \bar D^*$ in $I=1$, with quantum numbers $I^G(J^{PC})= 1^-(2^{++})$ with a mass around 3920 MeV and a width of about 150 MeV. The mass of the state can have large uncertainties since it is generated by the exchange of $J/\psi$ in the $t$-channel, which is suppressed by the large $J/\psi$ mass, such that other subleading mechanisms of the interaction could play some role. The $I=1$ only appears in the state  $J^{PC}=2^{++}$, since the tensor state is the one where the interaction is stronger. It is thus tempting to investigate whether the peak seen in the BESIII reaction \cite{besexp} can be interpreted in terms of such a state below the $D^* \bar D^*$ threshold. This is one of the aims of the present paper, but simultaneously we shall also investigate other possibilities, like the one suggested in the experimental paper \cite{besexp} of a $J^P=1^+$ state, or even the result of using simply phase space in $D$-waves. 

  One element of information that makes this investigation opportune is the fact that if the state were a $J^P=1^+$ produced in $S$-wave, as assumed in the experimental work \cite{besexp}, it easily decays into $\pi J/\psi$ exchanging a $D$ meson in the $t$-channel. This is also the decay channel of the $Z_c(3900)$ reported in Ref.~\cite{zc3900}, which would then have the same quantum numbers as the state claimed in Ref.~\cite{besexp}. However, while a peak is clearly seen in the 
$\pi J/\psi$ invariant mass distribution for the case of the $Z_c(3900)$, no trace of a peak is seen around 4025 MeV (see Fig.~4 of Ref.~\cite{zc3900}) in spite of using the same reaction and the same $e^+ e^-$ energy. 
  
  The argument exposed above gets extra support when one digs a bit more on the reaction used. The energy of 4260 MeV for the $e^+ e^-$ pair is not chosen arbitrarily but it corresponds to the energy of the X(4260) discovered in  Ref.~\cite{Aubert:2005rm} and later on reported in Ref.~\cite{He:2006kg,Yuan:2007sj,Coan:2006rv,Lees:2012cn}. This resonance was interpreted in Ref.~\cite{jpsikkbar} as a molecular state of $J/\psi K \bar K$, in analogy to the $\phi(2160)$, which was interpreted as a $\phi K \bar K$ in 
Ref.~\cite{phikkbar}. It has also been suggested a $D_1\bar D$ bound state~\cite{arx1}, because of the proximity of the threshold to the mass of the state, or as a hadrocharmonium state~\cite{arx2} (see also a discussion in Ref.~\cite{arx3}). Actually, the structure obtained in Ref.~\cite{jpsikkbar} for the X(4260) was of a $J/\psi$, together with a cluster of $K \bar K$ forming the $f_0(980)$. In the decay of this state the $f_0(980)$ goes to $\pi \pi$. It is quite interesting to see that in a posterior experiment~\cite{zc3900} the Dalitz plot for the decay of the X(4260) into $J/\psi \pi \pi$, as well as the mass distributions, shows a clear correlation of the $\pi \pi$ pair around the  $f_0(980)$ mass. The $J/\psi, ~K,~ \bar K$ states in the molecular state of the X(4260) are all in $S$-wave in Ref.~\cite{jpsikkbar}, hence the molecule contains an $S$-wave  $J/\psi \pi$ component to start with, that upon interaction can lead to $S$-wave resonances like the $Z_c(3900)$, and the  $Z_c(4025)$ if the quantum numbers are those assumed in Ref.~\cite{besexp}. Yet, as mentioned before, there is no trace of a peak around 4025 MeV in the invariant mass of $J/\psi \pi$. 

With respect to quantum numbers it is worth noting that the $e^+ e^-$ does not have a defined $I^G$. Even if one produces the X(4260), this resonance does not have $I^G$ defined in the PDG \cite{pdg}, and even if one assumes it to be the $I^G=0^-$ of Ref.~\cite{jpsikkbar}, one must note that in the $e^+ e^-$ reaction at that energy not only this broad resonance is produced but it is accompanied by a background with equal strength \cite{Aubert:2005rm}. Thus, G parity is a quantum number not to worry about in the reaction. The isospin of the $D^* \bar D^*$ state, however, is determined from the charged nature of the detected $D^* \bar D^*$ state. For this same reason one does not have to consider C parity. Hence, $I(J^P)$ are the basic quantum numbers to worry about. 

   There is another element worth considering with respect to associating a new resonance to the $D^* \bar D^*$ mass spectrum of Ref.~\cite{besexp}. It was found in 
Ref.~\cite{junkojuan} that a single channel with an energy independent potential can generate bound states but not resonance states above threshold. In the local hidden gauge approach the potentials in the heavy quark sector are practically energy independent and, furthermore, the mixing of channels is small, corresponding to subleading terms in the heavy quark mass ($m_Q$) counting, and even smaller than the diagonal terms for $I=1$, which are also suppressed in that counting \cite{xiaojuan}. One should note, however, that in works like \cite{HidalgoDuque:2012pq,Nieves:2012tt} which use only the restrictions of  Heavy Quark Spin Symmetry (HQSS) but no specific dynamics, mixing of channels is allowed. Yet, from the perspective of the dynamics of the local hidden gauge extrapolations used in the heavy quark sector, a resonant state of $D^* \bar D^*$ above threshold is not easy to support.

   There are further considerations to make. Indeed, the threshold energy for  $\pi D^* \bar D^*$ is 4160 MeV, which is only 100 MeV below the energy of the 
$e^+ e^-$ initial state. There are only 100 MeV of phase space in a reaction with three particles in the final state, where the invariant mass distribution for $D^* \bar D^*$ goes to zero at the beginning and the end of the phase space. This by itself creates a bump narrower than 100 MeV, in particular if the reaction occurs in $D$-waves. 

   Nevertheless, as usual when a claim is made of a new resonance, theoreticians dig into possible answers for the structure of the new state. In this sense it is worth quoting theoretical work done in this direction. In Ref.~\cite{guojuan} HQSS is used to make predictions for states containing one $D$ or $D^*$ and one $\bar D$ or  $\bar D^*$. Assuming the  X(3872) and $Z_b(10610)$ to be $D \bar D^*$ and $B \bar B^*$  molecules, the authors predict a series of new
hadronic molecules, including the $Z_c(3900)$ and the $Z_c(4025)$. The latter ones would correspond to bound states (with uncertainties of about 50 MeV in the binding) of $D \bar D^* + c.c.$ and $D^* \bar D^*$ respectively, with quantum numbers $I(J^{PC})=1(1^{+-})$. It is to be stressed that the states, even with uncertainties, always come in the bound region. In Refs.~\cite{Chen:2013omd, Cui:2013vfa}, using QCD sum rules and assuming a structure of $D^* \bar D^*$, the authors obtain a possible $I(J^{P})=1(1^+)$ state compatible with the $Z_c(4025)$ albeit with around $\pm 250$ MeV uncertainty in the energy. Recently, a study of the $D^*\bar D^*$ system has also been done within QCD sum rules, projecting the correlation function on spin-parity $0^+$, $1^+$ and $2^+$, and in the three cases a state with mass $ 3950\pm100$ MeV is found~\cite{soon}. The central value of the mass of these states is more in line with the results of Ref.~\cite{raquelxyz}, although with the error bar,
they could also be related to a resonance. In Ref.~\cite{He:2013nwa} the $Z_c$ state is looked up from a different perspective and, using pion exchange, a $D^* \bar D^*$ state with $I(J^{P})=1(1^+)$ compatible with the $Z_c(4025)$ is obtained. One should note that the input used in this latter work is quite different to the one in Ref.~\cite{guojuan} since in HQSS the pion exchange is subdominant \cite{xiaojuan}. One should also note that for $I=1$ states, there is a cancellation for equal masses (or large momenta) of $\pi, \eta, \eta'$ between the exchange of these pseudoscalars, as discussed in Ref.~\cite{xiaoaltug}. Finally, in Ref.~\cite{Qiao:2013dda}, using a tetraquark structure and QCD sum rules, a state with $I(J^{P})=1(2^+)$
compatible with $Z_c(4025)$ is obtained, once again with a large error in the energy of 190 MeV. These would be the quantum numbers of the state predicted in 
Ref.~\cite{raquelxyz}. A different idea is exposed in Ref.~\cite{Wang:2013qwa}, where a pion and the $D^* \bar D^*$ state are produced from the X(4260) and the $D^* \bar D^*$ state is left to interact with the pion being a spectator (initial single-pion emission mechanism). Although it is not mentioned whether the $D^* \bar D^*$ interaction produces a resonance with certain quantum numbers, as in Ref.~\cite{raquelxyz}, the authors show that the mechanism can produce some enhancement in the $D^* \bar D^*$ invariant mass distribution just above threshold.

As we can see, the theoretical panorama is rather diverse and uncertainties too large to serve as a support for the experimental claim of the $Z_c(4025)$. In the present paper we shall take a different path and investigate different ways in which the experimental data can be reproduced. The fact that we shall be able to describe the data with different options will render the claims for the $Z_c(4025)$ weak. Together with the former arguments about the difficulty to get a resonant state of two heavy mesons above threshold, the case for the  $Z_c(4025)$ with that mass is further weakened, but the peak observed being a consequence of a $D^* \bar D^*$ state below threshold remains a clear possibility. 
\section{Formalism}    
\subsection{$D$-wave production mechanism for a $J^{P}=2^+$ state}
\begin{figure}
\includegraphics[width=0.7\textwidth]{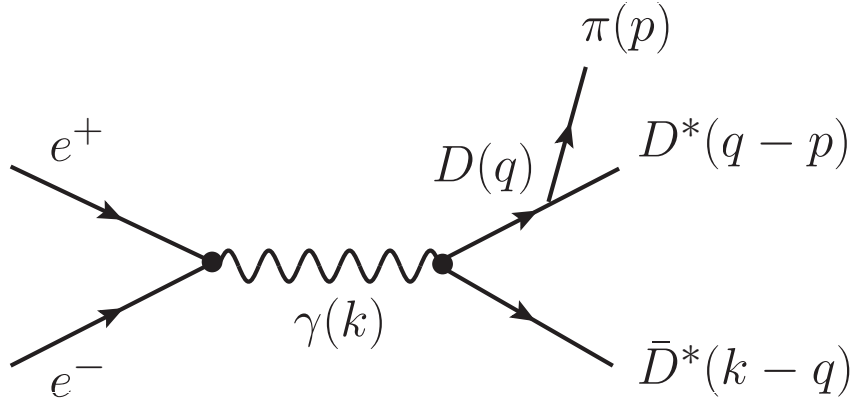}
\caption{Feynman diagram of a mechanism allowing the production of $D^*\bar{D}^*$ in $D$-waves. Momenta are shown in brackets.}\label{react}
\end{figure}

The production of $D^*\bar{D}^*$ in $D$-waves is one of the options that we shall exploit here, keeping in mind the possibility that the peak seen in Ref.~\cite{besexp} could be the $1^-(2^{++})$ state predicted in Ref.~\cite{raquelxyz} below the $D^*\bar{D}^*$ threshold. We first look at a mechanism that reassures us that the production of this $2^{++}$ resonance is possible in this reaction. Several mechanisms could be thought of, but it suffices to look at the one of Fig.~\ref{react}.
The structure of the amplitude for this diagram is given by
\begin{align}
t\propto L_\nu\epsilon^{\mu\nu\alpha\beta}k_\mu q_\alpha\frac{1}{q^2-m^2_D+i\epsilon}\epsilon_\beta(\bar{D}^*)\epsilon_\delta(D^*) (q+p)^\delta,\label{tprop}
\end{align}
where $p$, $k$, $q$ are the momenta depicted in Fig.~\ref{react} and
\begin{align}
L_\nu\equiv\bar{v}(p^+)\gamma_\nu u(p^-).\label{L}
\end{align}
In Eq.~(\ref{L}),  $p^+$ ($p^-$) represents the momentum of the $e^+$ ($e^-$). To determine the amplitude of Eq.~(\ref{react}), it is convenient  to work in the $e^+ e^-$ center of mass (CM) frame, in which the three momentum of the photon, $\vec{k}$, is zero, leaving in this way only the contribution from the $k^0$ component in Eq.~(\ref{react}). On the other hand, in the reaction depicted in Fig.~\ref{react}, the external three momenta of the $D^*$ and $\bar{D}^*$ are small, a fact which allows us to drop the $\epsilon_0$ component of the polarization vectors, as done in Ref.~\cite{raquelxyz}. Thus, Eq.~(\ref{tprop}) gets simplified to,
\begin{align}
t\propto k^0\epsilon^{ijk}L_i q_j(q+p)_m\frac{1}{q^2-m^2_D+i\epsilon}\epsilon_k(\bar{D}^*)\epsilon_m(D^*).\label{tprop2}
\end{align} 
As can be seen, Eq.~(\ref{tprop2}) contains the term $q_j q_m$, which  carries $D$-wave. On the other hand it is also interesting to see that the combination $\epsilon_m(D^*)\epsilon_k(\bar{D}^*)$ contains spin $S=2$ for the vector mesons. Indeed, for low momenta of the vectors the spin projectors over spin 0, 1, 2 are given by~\cite{raquelvec}:
\begin{align}
P^{(0)}&=\frac{1}{3}\vec{\epsilon}\cdot\vec{\epsilon}^{\,\prime}\delta_{km},\nonumber\\
P^{(1)}&=\frac{1}{2}\left(\epsilon_m\epsilon^\prime_k-\epsilon_k\epsilon^\prime_m\right),\label{proj}\\
P^{(2)}&=\frac{1}{2}\left(\epsilon_m\epsilon^\prime_k+\epsilon_k\epsilon^\prime_m\right)-\frac{1}{3}\vec{\epsilon}\cdot\vec{\epsilon}^{\,\prime}\delta_{km},\nonumber
\end{align}
where $\vec{\epsilon}$ is the polarization vector of the $D^*$ and $\epsilon^\prime$ the one of the  $\bar{D}^*$. It is easy to see that,
\begin{align}
\epsilon_m\epsilon^\prime_k=P^{(0)}+P^{(1)}+P^{(2)}
\end{align}
and, hence, the amplitude of Eq.~(\ref{tprop2}) has components of $D$-wave and $S=2$ for the two vector mesons.

\begin{figure}
\includegraphics[width=0.7\textwidth]{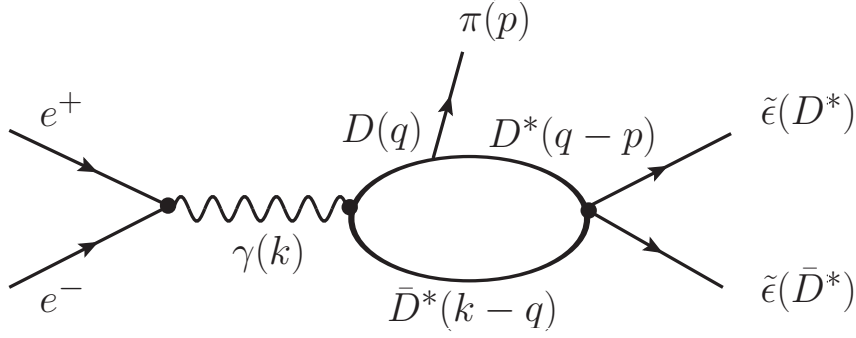}
\caption{Loop function accounting for the interaction of the two vector mesons}\label{react2}
\end{figure}

It remains to see that the $D$-wave structure and the $S=2$ character of the $D^*\bar{D}^*$ system are preserved upon interaction of the final $D^*\bar{D}^*$ states to produce the state predicted in Ref.~\cite{raquelxyz}. This process is depicted diagrammatically in Fig.~\ref{react2}. In this case, the amplitude is given by

\begin{align}
t^{(\textrm{loop})}\propto& \int \frac{d^4 q}{(2\pi)^4}\frac{1}{q^2-m^2_D+i\epsilon}\frac{1}{(q-p)^2-m^2_{D^*}+i\epsilon}\frac{1}{(k-q)^2-m^2_{\bar{D}^*}+i\epsilon}\nonumber\\
&\times k^0\epsilon^{ijk}L_i q_j (q+p)_m\epsilon_m(D^*)\epsilon^\prime_k(\bar{D}^*)t_\textrm{VV}\left\{\frac{1}{2}\left(\epsilon_{i^\prime}\epsilon^\prime_{j^\prime}+\epsilon_{j^\prime}\epsilon^\prime_{i^\prime}\right)-\frac{1}{3}\vec{\epsilon}\cdot\vec{\epsilon}^{\,\prime}\delta_{i^\prime j^\prime}\right\}\nonumber\\
&\times\left\{\frac{1}{2}\left(\tilde{\epsilon}_{i^\prime}\tilde{\epsilon}^\prime_{j^\prime}+\tilde{\epsilon}_{j^\prime}\tilde{\epsilon}^\prime_{i^\prime}\right)-\frac{1}{3}\vec{\tilde{\epsilon}}\cdot\vec{\tilde{\epsilon}}^{\,\prime}\delta_{i^\prime j^\prime}\right\}.\label{integ}
\end{align}
Since $\vec{k}=0$, the integral in Eq.~(\ref{integ}),  ignoring the polarization vectors, can only provide a structure like
\begin{align}
A\delta_{jm}+Bp_jp_m,\label{str}
\end{align}
in terms of the pion momentum. We can see again the $D$-wave character in the second term of Eq.~(\ref{str}). As to the spin structure, after the contraction of the space-like vector polarizations,
\begin{align}
\sum_\textrm{pol}\epsilon_m\epsilon_{i^\prime}=\delta_{m i^\prime},
\end{align}
we obtain  the following combination in term of the polarization of the external vector mesons, $\tilde{\epsilon}$:
\begin{align}
\frac{1}{2}\left(\tilde{\epsilon}_m\tilde{\epsilon}^\prime_k+\tilde{\epsilon}_k\tilde{\epsilon}^\prime_m\right)-\frac{1}{3}\vec{\tilde{\epsilon}}\cdot\vec{\tilde{\epsilon}}^{\,\prime}\delta_{mk}.\label{comb}
\end{align}
Comparing Eq.~(\ref{comb}) with the spin 2 projector of Eq.~(\ref{proj}), one can see that the spin $S=2$ structure of the pair of vector mesons is preserved through the interaction. Note that the loop of Fig.~\ref{react2} is intrinsically different from the one of Ref.~\cite{Wang:2013qwa} where the initial single-pion emission mechanism has the pion line going out from the same $\gamma D^*\bar{D}^*$ vertex in Fig.~\ref{react2} ($X(4260) D^*\bar{D}^*\pi$ vertex in Ref.~\cite{Wang:2013qwa}). Here the pion is emitted from inside the loop and is essential to provide the desired $D$-wave behavior. The structure of the vertices in Ref.~\cite{Wang:2013qwa} allows $S$-wave formation, hence, leading to $D^*\bar{D}^*$ in $J^P=1^+$.
\subsection{Invariant mass distribution}
In the production of a $D^*\bar{D}^*$ resonant state we shall assume that the amplitude depends on the invariant mass of $D^*\bar{D}^*$, $M_{D^*\bar{D}^*}$, as done in Ref.~\cite{besexp}. In this case the differential cross section is given by~\cite{danibump}
\begin{align}
\frac{d\sigma}{d M_{D^*\bar{D}^*}}\propto \frac{m^2_e}{s\sqrt{s}}p\tilde{q}\left|T\right|^2F_L,
\end{align}
where $p$ is the pion momentum in the $e^+ e^-$ CM frame and $\tilde{q}$ is the $D^*$ momentum in the $D^*\bar{D}^*$ CM frame:
\begin{align}
p&=\frac{\lambda^{1/2}(s,m^2_\pi,M^2_{D^*\bar{D}^*})}{2\sqrt{s}},\\
\tilde{q}&=\frac{\lambda^{1/2}(M^2_{D^*\bar{D}^*},m^2_{D^*},m^2_{\bar D^*})}{2M_{D^*\bar{D}^*}}.
\end{align}
The factor $F_L=p^{2L}$ is needed to account for the partial wave in which the $D^*\bar{D}^*$ system is produced ($L=0$ for $S$-waves and $L=2$ for $D$-waves), and $T$ is an amplitude which we parametrize as
\begin{align}
T=\frac{A}{M^2_{D^*\bar{D}^*}-M^2_R+iM_R\Gamma_R},\quad A\equiv\textrm{constant}
\end{align}
for the case of a resonance produced with mass $M_R$ and width  $\Gamma_R$. In case of a pure phase space, $T$ would be taken as a constant. 

In general, as done in Ref.~\cite{besexp}, the $D^*\bar D^*$ invariant mass distribution can have contribution from a small background proportional to the phase space, and from combinatorial backgrounds (estimated by combining a reconstructed $D^+$ with a pion of the wrong charge, see Ref.~\cite{besexp} for more details) referred to as wrong sign (WS) background. Thus, we can write
\begin{align}
\frac{d\sigma}{d M_{D^*\bar{D}^*}}=\frac{m^2_e}{s\sqrt{s}}p\tilde{q}\left(\left|T\right|^2F_L+B\right)+\textrm{WS},\label{cross}
\end{align}
where $B$ represents a constant.
\section{results}
\subsection{Invariant mass distribution}\label{res}
In this section we show the results found using Eq.~(\ref{cross}) for the $D^*\bar D^*$ invariant mass distribution of the process $e^+ e^- \to (D^* \bar D^*)^{\pm} \pi^{\mp}$  at the same energy than the one considered in Ref.~\cite{besexp}, that is, $\sqrt{s}=4.26$ GeV. We determine the $D^*\bar D^*$ spectrum for the following cases:

\begin{enumerate}[I.]
\item Production of a $1^+$ $D^*\bar D^*$ resonance in relative $S$-wave with the pion.
\item Production of a $1^+$ $D^*\bar D^*$ bound state in relative $S$-wave with the pion.
\item Production of a $2^+$ $D^*\bar D^*$ resonance in relative $D$-wave with the pion.
\item Production of a $2^+$ $D^*\bar D^*$  bound state in relative $D$-wave with the pion
\item A pure $D$-wave background.
\end{enumerate}

As we will show in the next sections, there is more than one option compatible with the invariant mass distribution obtained by the BES collaboration in Ref.~\cite{besexp}.

While determining Eq.~(\ref{cross}), we have four unknown parameters: the resonance mass, $M_R$, its width, $\Gamma_R$,  the magnitude of the resonant amplitude ($A$) and that of the
phase space background ($B$). The strategy followed to constrain these parameters consists of performing a fit to the experimental data with the mentioned degrees of freedom. In the fitting process we also take into account
the same wrong sign events background (WS) determined by the BES collaboration in the study of the reaction $e^+ e^- \to (D^* \bar D^*)^{\pm} \pi^{\mp}$~\cite{besexp}. 
We minimize the $\chi^2$ and consider the result whose $\chi^2$ per degrees of freedom (n.d.o.f) is $\sim$1 as a good fit. This is the same criteria as the one of Ref.~\cite{besexp}, in which a fit to the data is performed
using the option I listed above and a solution compatible with the data is obtained with $\chi^2/\textrm{n.d.o.f}=0.92$. 
\subsection*{Case I: A $1^+$ $D^*\bar D^*$  resonance in relative $S$-wave with the pion.}\label{cassabove}
\begin{figure}
\includegraphics[width=0.5\textwidth]{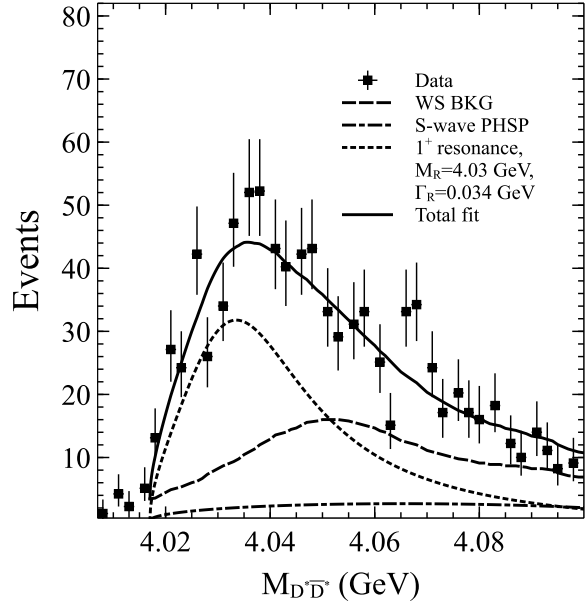}
\caption{Invariant $D^*\bar D^*$ mass distribution obtained using Eq.~(\ref{cross}) for the case of a $1^+$ resonance in relative $S$-wave with respect to pion, mass $M_R=4030$ MeV and width $\Gamma_R=34$ MeV. From now onwards we use the following nomenclature in the figures:  PHSP means phase space, BKG means background and WS means wrong sign.} \label{sabove}
\end{figure}

In Fig.~\ref{sabove} we show the $D^*\bar D^*$ invariant mass distribution obtained from a fit to the data considering a resonance in relative $S$-wave with the pion,  mass $4030$ MeV and a narrow width, $34$ MeV. These values are similar to those adopted by the BES collaboration to explain the data~\cite{besexp}. As can be seen, the agreement with the data is good and the $\chi^2/\textrm{n.d.o.f}\sim 1.09$. 
This result could indicate the existence of a resonance in relative $S$-wave, around 10-13 MeV above the $D^*\bar D^*$ threshold and narrow width $\sim 30$ MeV. However, as we are going to see in the next sections, this is not the only possibility which can satisfactorily explain the data. 

Note that the formation of the $1^+$ state is  possible with $L=0$ and $L=2$. We take the common choice of associating the lowest possible $L$, that normally dominates the process.

\subsection*{Case II: A $1^+$ $D^*\bar D^*$  bound state in relative $S$-wave with the pion.}
\begin{figure}[h!]
\includegraphics[width=0.5\textwidth]{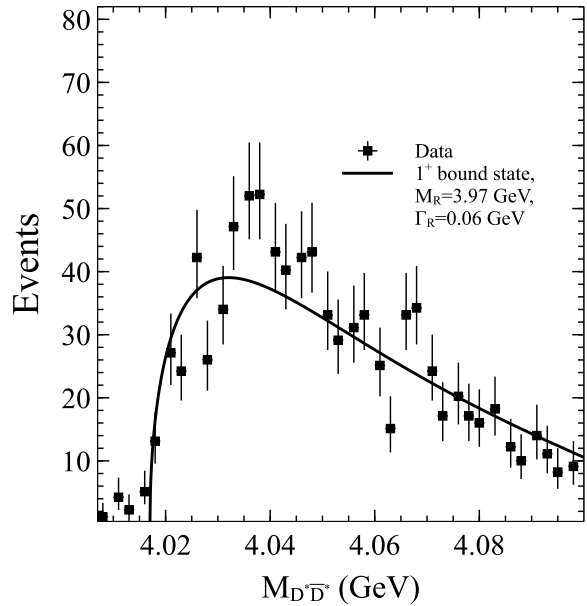}
\caption{Invariant mass $D^*\bar D^*$ spectrum obtained using Eq.~(\ref{cross}) to fit the data of Ref.~\cite{besexp}. In this case, we consider the formation of a $1^+$ $D^*\bar D^*$ bound state in relative $S$-wave with the pion.} \label{sbelow}
\end{figure}

As has been found in several works~\cite{danibump,MartinezTorres:2012du,miguel}, states below the threshold of a process can generate bumps close to threshold, leading to claims of resonance above the threshold of the reaction studied and, thus, to a misinterpretation of the observed signal. It would be then interesting to test if a $1^+$ state in relative $S$ wave with the pion, as assumed by the BES collaboration, but below the $D^*\bar D^*$ threshold could explain the results found in Ref.~\cite{besexp}. A best fit is obtained for a mass $M_R\sim 3970$ MeV and width $\Gamma_R\sim 60$ MeV. As can be seen in Fig.~\ref{sbelow} this option is less suitable ($\chi^2/\textrm{n.d.o.f}\sim 1.4$) than the previous one to explain the data. Note that the result shown in Fig.~\ref{sbelow} correspond only to the resonance signal, without any kind of background. The inclusion of a background
further worsens the agreement with the data and, hence, $\chi^2/\textrm{n.d.o.f}$ is much larger than 1.4.
\subsection*{Case III: A $2^+$ $D^*\bar D^*$ bound state in relative $D$-wave with the pion.}\label{sec3}
\begin{figure}[h!]
\includegraphics[width=0.5\textwidth]{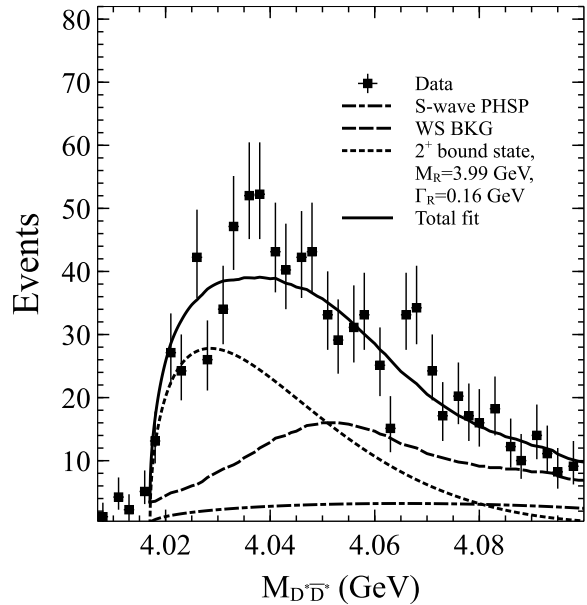}
\caption{Invariant mass distribution obtained from Eq.~(\ref{cross}) for the case of a $2^+$ $D^*\bar D^*$ bound state in $D$-wave with the pion. The best fit to the data gives a mass $M_R=3990$ MeV and width $\Gamma_R=160$ MeV.} \label{dbelow}
\end{figure}

In Ref.~\cite{raquelxyz} a $D^*\bar D^*$ bound state is found with isospin 1, spin-parity $2^+$, mass around 3900-3970 MeV and width $140-200$ MeV. The BES collaboration has attributed the signal observed in the $D^*\bar D^*$ invariant mass distribution of the process  $e^+ e^- \to (D^* \bar D^*)^{\pm} \pi^{\mp}$ to a $1^+$ resonance produced in relative $S$-wave with the pion. However, the spin-parity conservation does not rule out  a $D^*\bar D^*$ state with spin-parity $2^+$ (as the one found in Ref.~\cite{raquelxyz}) and relative angular momentum $L=2$, i.e., $D$-wave production. We show the $D^*\bar D^*$ spectrum calculated for this option in Fig.~\ref{dbelow}.  As can be seen, a bound state with a mass of $M_R=3990$ MeV and width $\Gamma_R=160$ MeV, although being below the $D^*\bar D^*$ threshold and broad, when convoluted with the phase space factor produces a narrower bump $10-15$ MeV above the $D^*\bar D^*$ threshold. The 
$\chi^2/\textrm{n.d.o.f}$ obtained in this case is $\sim 1.2$, thus this option is a plausible explanation for the signal found in Ref.~\cite{besexp}.

\subsection*{Case IV: A $2^+$ $D^*\bar D^*$ resonance in relative $D$-wave with the pion.}
\begin{figure}[h!]
\includegraphics[width=0.5\textwidth]{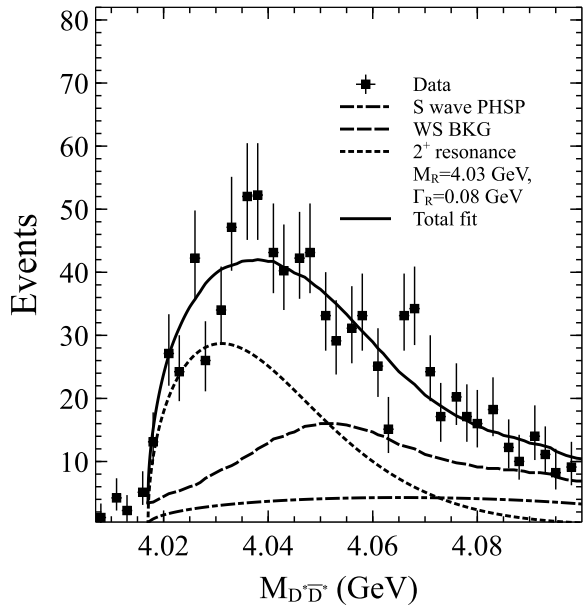}
\caption{Invariant mass distribution obtained considering Eq.~(\ref{cross}) to fit the data with a $2^+$ resonance in relative $D$-wave with the pion. We obtain from the fit: mass $M_R=4030$ MeV and width $\Gamma_R=80$ MeV.} \label{dabove}
\end{figure}

One could also consider the formation of a $2^+$ resonance in the $D^*\bar D^*$ system in relative $D$-wave, instead of a bound state as in case III, as the mechanism responsible for the spectrum obtained in Ref.~\cite{besexp}. The result is shown in Fig.~\ref{dabove} for the case  $M_R=4030$ MeV and width $\Gamma_R=80$ MeV. As can be seen, the solution found in this way is perfectly compatible with the data ($\chi^2/\textrm{n.d.o.f}\sim 1.1$). 
\subsection*{Case V: $D$-wave background.}
\begin{figure}[h!]
\includegraphics[width=0.5\textwidth]{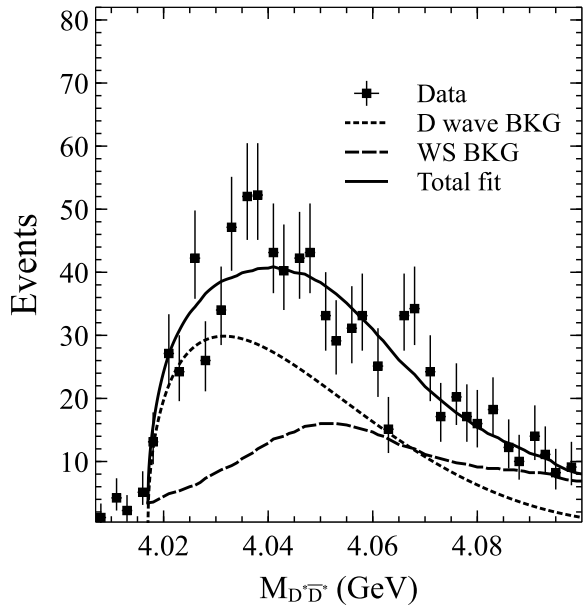}
\caption{Invariant mass spectrum obtained from a fit to the data with a pure $D$-wave background.} \label{dbackg}
\end{figure}

One could also wonder if a resonance or bound state is needed to explain the signal found by the BES collaboration in Ref.~\cite{besexp}. In Fig.~\ref{dbackg} we show the solution obtained considering a pure $D$-wave background in Eq.~(\ref{cross}). Surprisingly, this possibility is also compatible with the data, obtaining a $\chi^2/\textrm{n.d.o.f}\sim 1.1$.

As a summary of this section, we have studied the origin of the signal found $10-15$ MeV above the $D^*\bar D^*$ threshold in Ref.~\cite{besexp}. We conclude that, although the existence of a $1^+$ resonance with mass $\sim 4030$ MeV, narrow width, $\sim 30$ MeV, and relative $S$-wave  with respect the pion, as assumed in Ref.~\cite{besexp}, is compatible with the data, there are more options with which the signal can be explained: a broad $2^+$ bound $D^*\bar D^*$ state in relative $D$-wave with the pion of the reaction considered; a $2^+$ resonance above the $D^*\bar D^*$ threshold in $D$-wave with the pion or simply a $D$-wave background. All these options are equally plausible to describe the spectrum and the signal found in Ref.~\cite{besexp}.

\subsection{Energy dependence of the $D^*\bar D^*$ invariant mass distribution}
It would be interesting to know if there could be a way of finding which option, out of the different ones studied in Sec.~\ref{res} and compatible with the data of Ref.~\cite{besexp}, is responsible for the signal observed close to the $D^*\bar D^*$ threshold. A way to do this consists in investigating the dependence of the solutions found in Sec.~\ref{res} with the center of mass energy, $\sqrt{s}$. The experiment considered in Ref.~\cite{besexp} was studied at a center of mass energy of $\sqrt{s}=4.26$ GeV. In this section we show how the results of Sec.~\ref{res}  change when the center of mass energy is taken to be $\sqrt{s}=4.4$ GeV and $\sqrt{s}=4.6$ GeV. It should be added here that we have taken the background of WS events given in Ref.~\cite{besexp} for all values of $\sqrt{s}$, although it could also change with the center of mass energy.
\subsection*{Case I: A $1^+$ $D^*\bar D^*$  resonance in relative $S$-wave with the pion.}\label{saboenerg}
\begin{figure}
\includegraphics[width=0.5\textwidth]{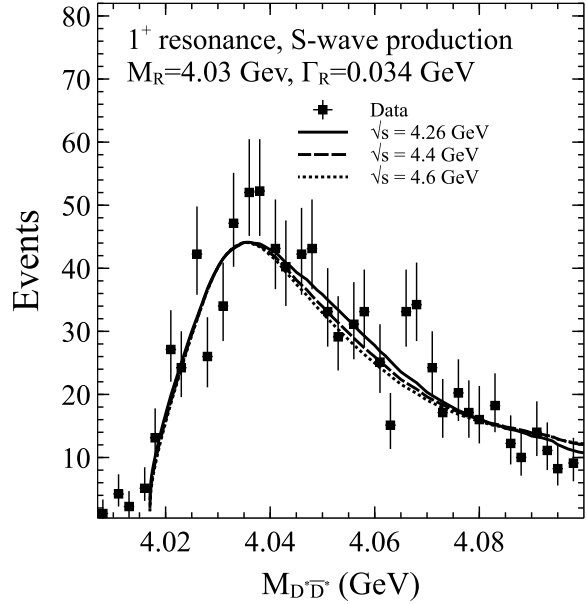}
\caption{Invariant mass distribution obtained from a fit to the data with a $1^+$ resonance in relative $S$-wave with the pion, mass $M_R=4030$ MeV and width $\Gamma_R=34$ MeV for different$\sqrt{s}$ values.} \label{saboveenerg}
\end{figure}

We show in Fig.~\ref{saboveenerg} the $D^*\bar D^*$ invariant mass distribution for the case of a $1^+$ resonance with 4030 MeV of mass and 34 MeV of width in relative $S$-wave with the pion for three values of $\sqrt{s}$, 4.26, 4.4 and 4.6 GeV. To compare them, we have renormalized (here and in the following cases) the results associated to the energies $\sqrt{s}=4.4$ GeV and $\sqrt{s}=4.6$ GeV to the one of $\sqrt{s}=4.26$ GeV.  As can be seen, not much changes in the $D^*\bar D^*$ invariant mass spectrum while varying  $\sqrt{s}$.
\subsection*{Case II: A $1^+$ $D^*\bar D^*$  bound state in relative $S$-wave with the pion.}
We do not consider this case, since the fit shown in Fig.~\ref{sbelow} and the $\chi^2/\textrm{n.d.o.f}$ obtained already indicate that this option is the least plausible one to explain the $D^*\bar D^*$ spectrum found in Ref.~\cite{besexp}.
\subsection*{Case III: A $2^+$ $D^*\bar D^*$ bound state in relative $D$-wave with the pion.}
\begin{figure}
\includegraphics[width=0.5\textwidth]{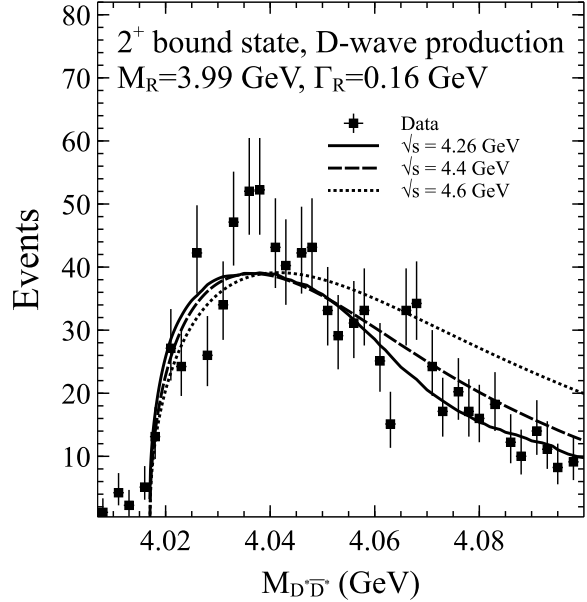}
\caption{Invariant mass distribution obtained for a $2^+$ $D^*\bar D^*$ bound state ($M_R=3990$ MeV, $\Gamma_R=160$ MeV) in relative $D$ wave with the pion for different values of $\sqrt{s}$.} \label{dbelowenerg}
\end{figure}

In case of production of a broad bound state at 3990 MeV in $D$-wave with the pion, the $D^*\bar D^*$ invariant mass distribution changes more than in case I when $\sqrt{s}$ is increased, specially
for $\sqrt{s}\sim4.6$ GeV (see Fig.~\ref{dbelowenerg}). The different energy behavior between the option tried in case I and the one considered here can be useful to determine if the signal is due to a resonance in relative $S$-wave or a bound state in relative $D$-wave with the pion present in the reaction studied in Ref.~\cite{besexp}.

\subsection*{Case IV: A $2^+$ $D^*\bar D^*$ resonance in relative $D$-wave with the pion.}
\begin{figure}
\includegraphics[width=0.5\textwidth]{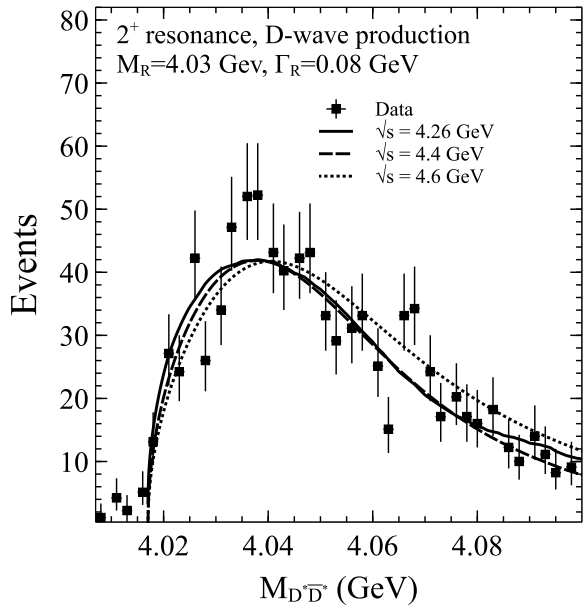}
\caption{Invariant mass distribution obtained for a $2^+$ $D^*\bar D^*$ resonance ($M_R=4030$ MeV, $\Gamma_R=80$ MeV) in relative $D$ wave with the pion for different values of $\sqrt{s}$.} \label{daboveenerg}
\end{figure}

In Fig.~\ref{daboveenerg} we show the results found for the $D^*\bar D^*$ spectrum while varying $\sqrt{s}$ for the case of a $2^+$ resonance in the $D^*\bar D^*$ system with mass 4030 MeV and width 80 MeV which is in relative $D$ wave with respect to the pion. As can be seen,
the energy dependence found is very weak, and, as in case I, the invariant mass distribution obtained for the three energies considered is compatible, within the error bars, with the data points found for $\sqrt{s}=4.26$ GeV by the BES collaboration in Ref.~\cite{besexp}. However, the changes obtained in the spectrum when varying $\sqrt{s}$ are more appreciable than in case of the $1^+$ resonance produced in $S$-wave (case I).
Thus, according to our findings, if an experimental study of the $D^*\bar D^*$ spectrum is done at different center of mass energies, it would be  possible to identify if the signal observed in the invariant mass distribution is due to the formation of resonance or a bound state in the $D^*\bar D^*$ system. However, it would be difficult to judge if the signal correspond to a $1^+$ resonance or to a $2^+$ resonance.

\subsection*{Case V: $D$-wave background.}
\begin{figure}[h!]
\includegraphics[width=0.5\textwidth]{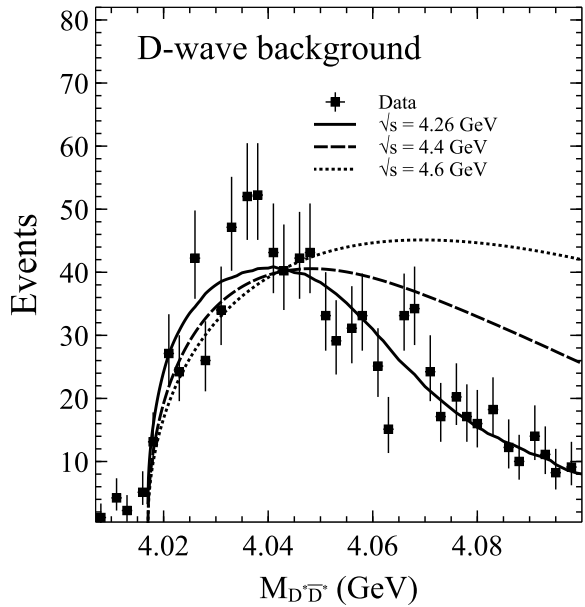}
\caption{Invariant mass distribution obtained with a $D$ wave background for different values of $\sqrt{s}$.} \label{dbackenerg}
\end{figure}

For this case, the changes observed in the $D^*\bar D^*$ invariant mass distribution while increasing $\sqrt{s}$ from 4.26 GeV to 4.6 GeV are very pronounced, a finding which definitely should be helpful in ruling out a $D$-wave background as responsible for the signal reported in Ref.~\cite{besexp} (see Fig.~\ref{dbackenerg}).
\section{Conclusions}
In this manuscript we have determined the $D^*\bar D^*$ invariant mass spectrum for the reaction $e^+ e^- \to (D^* \bar D^*)^{\pm} \pi^{\mp}$, studied by the BES collaboration~\cite{besexp}. 
We have found that, apart from the solution proposed in Ref.~\cite{besexp} to explain the signal observed close to the $D^*\bar D^*$ threshold, which is a $1^+$ resonance with mass $\sim$ 4030 MeV and
width $\sim$ 30 MeV, other options are also equally compatible with the data: a molecular $2^+$ $D^*\bar D^*$ bound state or resonance in relative $D$-wave with the pion or just a pure $D$-wave background are
options which can not be disregarded. With the idea of motivating further experimental studies which could clarify the origin of the signal obtained in Ref.~\cite{besexp}, we study the modification experienced by the $D^*\bar D^*$ invariant mass spectrum
when the center of mass energy is varied from $\sqrt{s}=4.26$ GeV to 4.6 GeV. As a result, we find that it is possible to clarify if the spectrum of Ref.~\cite{besexp} is due to a resonance, to a bound state or a pure $D$-wave background. However, if the origin turns out to be a resonance, then, it would be difficult to know if its spin-parity is $1^+$ or  $2^+$. This information should be certainly useful for further experimental analysis and can be used to shed some light on the intriguing signal found by the BES collaboration.
\section*{Acknowledgments}
The authors would like to thank the Brazilian funding agencies FAPESP and CNPq for the financial support.
E. Oset acknowledges the financial support provided by NAPQCD during his stay at the University of S\~ao Paulo.
This work is partly supported by the Spanish Ministerio de Econom\'\i a y Competitividad and
European FEDER funds under the contract number FIS2011-28853-C02-01 and  FIS2011-28853-C02-02, and the Generalitat
Valenciana in the program Prometeo, 2009/090. We acknowledge the support of the European
Community-Research Infrastructure Integrating Activity Study of Strongly Interacting Matter
(acronym HadronPhysics3, Grant Agreement n. 283286) under the Seventh Framework Programme of the EU.

\end{document}